\documentclass[twocolumn,aps,showpacs,preprintnumbers,amsmath,amssymb, superscriptaddress]{revtex4-1}
\pdfoutput=1

\usepackage{bm}
\usepackage{graphicx}
\usepackage{color}

\begin{document}

\title{Direct spectroscopic evidence for completely filled Cu $3d$ shell in BaCu$_2$As$_2$ and $\alpha$-BaCu$_2$Sb$_2$}

\author{S. F. Wu}
\affiliation{Beijing National Laboratory for Condensed Matter Physics, and Institute of Physics, Chinese Academy of Sciences, Beijing 100190, China}
\author{P. Richard}\email{p.richard@iphy.ac.cn}
\affiliation{Beijing National Laboratory for Condensed Matter Physics, and Institute of Physics, Chinese Academy of Sciences, Beijing 100190, China}
\affiliation{Collaborative Innovation Center of Quantum Matter, Beijing, China}
\author{A. van Roekeghem}
\affiliation{Beijing National Laboratory for Condensed Matter Physics, and Institute of Physics, Chinese Academy of Sciences, Beijing 100190, China}
\affiliation{Centre de Physique Th\'{e}orique, Ecole Polytechnique, CNRS-UMR7644, 91128 Palaiseau, France}
\author{S. M. Nie}
\affiliation{Beijing National Laboratory for Condensed Matter Physics, and Institute of Physics, Chinese Academy of Sciences, Beijing 100190, China}
\author{H. Miao}
\affiliation{Beijing National Laboratory for Condensed Matter Physics, and Institute of Physics, Chinese Academy of Sciences, Beijing 100190, China}
\author{N. Xu}
\affiliation{Swiss Light Source, Paul Scherrer Institut, CH-5232 Villigen PSI, Switzerland}
\author{T. Qian}
\affiliation{Beijing National Laboratory for Condensed Matter Physics, and Institute of Physics, Chinese Academy of Sciences, Beijing 100190, China}
\author{B. Saparov}\thanks{Present address: Department of Mechanical Engineering and Materials Science, Duke University, Durham, NC 27708, USA}
\affiliation{Materials Science and Technology Division, Oak Ridge National Laboratory, 1 Bethel Valley Road, Oak Ridge, TN 37831, USA}
\author{Z. Fang}
\affiliation{Beijing National Laboratory for Condensed Matter Physics, and Institute of Physics, Chinese Academy of Sciences, Beijing 100190, China}
\affiliation{Collaborative Innovation Center of Quantum Matter, Beijing, China}
\author{S. Biermann}
\affiliation{Centre de Physique Th\'{e}orique, Ecole Polytechnique, CNRS-UMR7644, 91128 Palaiseau, France}
\affiliation{Coll\`{e}ge de France, 11 place Marcelin Berthelot, 75005 Paris, France}
\affiliation{European Theoretical Synchrotron Facility (ETSF), Europe}
\author{Athena S. Sefat}
\affiliation{Materials Science and Technology Division, Oak Ridge National Laboratory, 1 Bethel Valley Road, Oak Ridge, TN 37831, USA}
\author{H. Ding}\email{dingh@iphy.ac.cn}
\affiliation{Beijing National Laboratory for Condensed Matter Physics, and Institute of Physics, Chinese Academy of Sciences, Beijing 100190, China}
\affiliation{Collaborative Innovation Center of Quantum Matter, Beijing, China}

\date{\today}

\begin{abstract}
We use angle-resolved photoemission spectroscopy to extract the band dispersion and the Fermi surface of BaCu$_2$As$_2$ and $\alpha$-BaCu$_2$Sb$_2$. While the Cu $3d$ bands in both materials are located around 3.5 eV below the Fermi level, the low-energy photoemission intensity mainly comes from As $4p$ states, suggesting a completely filled Cu $3d$ shell. The splitting of the As $3d$ core levels and the lack of pronounced three-dimensionality in the measured band structure of BaCu$_2$As$_2$ indicate a surface state likely induced by the cleavage of this material in the collapsed tetragonal phase, which is consistent with our observation of a Cu$^{+1}$ oxydation state. However, the observation of Cu states at similar energy in $\alpha$-BaCu$_2$Sb$_2$ without the pnictide-pnictide interlayer bonding characteristic of the collapsed tetragonal phase suggests that the short interlayer distance in BaCu$_2$As$_2$ follows from the stability of the Cu$^{+1}$ rather than the other way around. Our results confirm the prediction that BaCu$_2$As$_2$ is an $sp$ metal with weak electronic correlations.\end{abstract}



\maketitle

Even today, the cuprate superconductors constitute the family of unconventional superconductors exhibiting the highest critical temperatures. The parent compounds of the cuprates are strongly correlated materials with Cu in the $3d^9$ configuration (Cu$^{2+}$) bridged by oxygen. Chalcogen and pnictogen atoms also serve as bridges in the layered Fe-based superconductors, for which typical electronic band renormalization factors of 2-5 are usually found \cite{RichardRoPP2011}. Naively, one would expect that a Cu-based material crystallizing in the same crystal structure as a Fe-based superconductor should also show strong electronic correlations and possible superconductivity, thus motivating the study of such compounds. Not only superconductivity is not observed in BaCu$_2$As$_2$ \cite{Sefat2012BaCu2As2}, this material has been predicted to be an $sp$ metal with a filled $3d$ shell \cite{singh2009electronic}. Unfortunately, there is still no spectroscopic evidence supporting this scenario.

Here we report an angle-resolved photoemission spectroscopy study of the electronic band dispersion and the Fermi surface of BaCu$_2$As$_2$ and $\alpha$-BaCu$_2$Sb$_2$. Our photon energy-dependent study reveals that the Cu $3d$ states are located around 3.5 eV below the Fermi level ($E_F$), whereas the intensity around $E_F$ is mainly derived from As 4$p$ states. The experimental Fermi surfaces of both materials are very similar and qualitatively consistent with that of $\alpha$-BaCu$_2$Sb$_2$ derived from generalized gradient approximation (GGA) calculations. Except for the lack of three-dimensionality in BaCu$_2$As$_2$ associated with a surface state following the cleavage of samples in the collapsed tetragonal phase, the experimental band dispersion are also consistent with non-renormalized GGA calculations, suggesting the absence of strong electronic correlations. Our ARPES results indicate that BaCu$_2$As$_2$ and $\alpha$-BaCu$_2$Sb$_2$ have a fully filled Cu $3d$ shell and that the stability of this electronic configuration favors the collapsed tetragonal phase of BaCu$_2$As$_2$.

Large single-crystals of BaCu$_2$As$_2$ and $\alpha$-BaCu$_2$Sb$_2$ were grown by the self-flux method \cite{Sefat2012BaCu2As2}. ARPES measurements were performed at the PGM and APPLE-PGM beamlines of the Synchrotron Radiation Center (Wisconsin) equipped with a VG-Scienta R4000 analyzer and a SES 200 analyzer, respectively. The energy and angular resolutions for the angle-resolved data were set at 10-30 meV and 0.2$^\circ$, respectively. The samples were cleaved \emph{in situ} and measured at 20 K in a working vacuum better than $5 \times 10^{-11}$ Torr. In the following, we label the momentum values with respect to the 1 Cu/unit cell Brillouin zone, and use $c'=c/2$ as the distance between two Cu planes. Raman scattering measurements were performed using Argon-Krypton laser lines in a back-scattering micro-Raman configuration with a triple-grating spectrometer (Horiba Jobin Yvon T64000) equipped with a nitrogen-cooled CCD camera. In this manuscript, we define $x$ and $y$ as the directions along the $a$ or $b$ axes, oriented 45$^{\circ}$ from the Cu-Cu bounds, while $x'$ and $y'$ are defined as the Cu-Cu directions. We performed first-principles calculations of the electronic band structure by using the full-potential linearized-augmented plane-wave (FP-LAPW) method implemented in the WIEN2K package for the crystal structures of BaCu$_2$As$_2$ and $\alpha$-BaCu$_2$Sb$_2$ \cite{Sefat2012BaCu2As2}. The exchange-correlation potential was treated using the generalized gradient approximation (GGA) based on the Perdew-Burke-Ernzerhof (PBE) approach \cite{Perdew_PRL77}. 

\begin{figure}[!t]
\begin{center}
\includegraphics[width=3.4in]{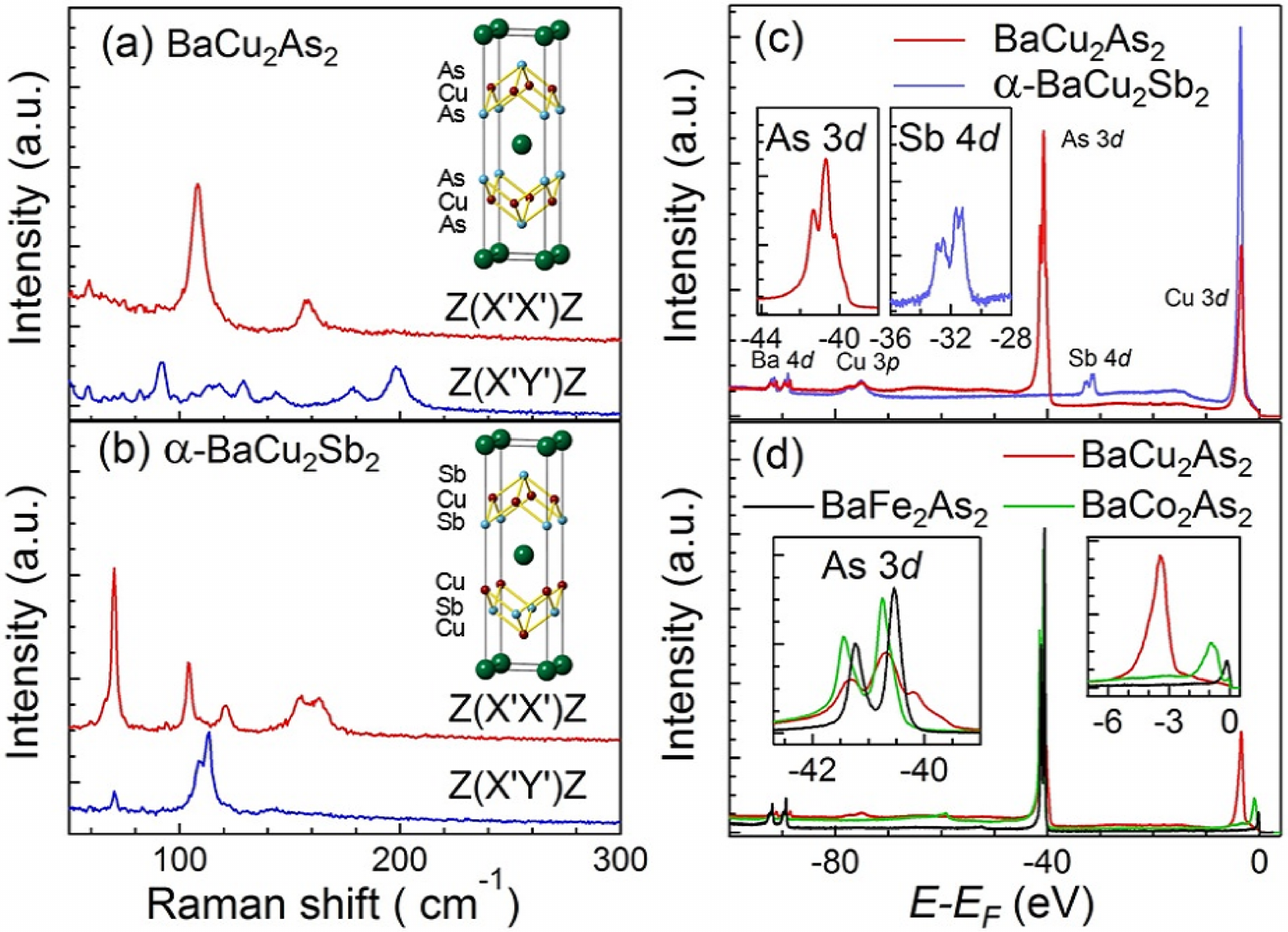}
\end{center}
\caption{\label{Fig1_core}(Color online) (a) and (b) Polarized $ab$ plane Raman spectra of BaCu$_2$As$_2$ and $\alpha$-BaCu$_2$Sb$_2$ at 294 K. Insets: corresponding crystal structures. (c) Core levels of BaCu$_2$As$_2$ and $\alpha$-BaCu$_2$Sb$_2$ recorded at 180 eV. The inset zooms on the As $3d$ and Sb $4d$ core levels of BaCu$_2$As$_2$ and $\alpha$-BaCu$_2$Sb$_2$, respectively. (d) Comparisons of the core levels of BaM$_2$As$_2$ for M = Fe, Co and Cu, recorded at 180 eV. The insets show the zooms on the As $3d$ core levels and the valence band.}
\end{figure}

The crystal structures of BaCu$_2$As$_2$ and $\alpha$-BaCu$_2$Sb$_2$ are closely related, as illustrated by the insets of Figs. \ref{Fig1_core}(a) and \ref{Fig1_core}(b).  While BaCu$_2$As$_2$ has the same crystal structure as BaFe$_2$As$_2$, described by the space group D$_{4h}^{17}$ (I4/mmm), $\alpha$-BaCu$_2$Sb$_2$, characterized by the space group D$_{4h}^{7}$(P4/nmm), has the relative positions of the Sb and Cu atoms exchanged in two successive layers. To confirm that our samples of BaCu$_2$As$_2$ and $\alpha$-BaCu$_2$Sb$_2$ have different crystal structures, we recorded Raman spectra at room temperature with incident and scattered light polarized in the $ab$ plane. Our analysis predicts 4 Raman active modes in BaCu$_2$As$_2$ (A$_{1g}$+B$_{1g}$+2E$_g$) and 9 in $\alpha$-BaCu$_2$Sb$_2$ (3A$_{1g}$+2B$_{1g}$+4E$_g$). While the in-plane polarization used does not allow us to detect E$_g$ modes, the A$_{1g}$ (related to the vibration of the pnictogens along the $c$-axis) and B$_{1g}$ (related to the vibration of Cu) channels are accessible in the $z(x'x')\bar{z}$ and $z(x'y')\bar{z}$ configurations, respectively. The experimental results for BaCu$_2$As$_2$ and $\alpha$-BaCu$_2$Sb$_2$ are shown in Figs. \ref{Fig1_core}(a) and \ref{Fig1_core}(b), respectively. We observe an A$_{1g}$ mode at 108.9 cm$^{-1}$ and a B$_{1g}$ mode at 198.2 cm$^{-1}$ in BaCu$_2$As$_2$, whereas 3 A$_{1g}$ modes at 70.4 cm$^{-1}$, 104 cm$^{-1}$ and 121.4 cm$^{-1}$ and 2 B$_{1g}$ modes at 108.7 cm$^{-1}$ and 113.4 cm$^{-1}$ are detected in $\alpha$-BaCu$_2$Sb$_2$. Although additional weak and unidentified peaks are also detected, the spectra of our BaCu$_2$As$_2$ and $\alpha$-BaCu$_2$Sb$_2$ samples are clearly different, thus proving that they correspond to different crystal phases.

We then confirm the elemental composition of our samples by showing the core level spectra of BaCu$_2$As$_2$ and $\alpha$-BaCu$_2$Sb$_2$ in Figs. \ref{Fig1_core}(c). In both materials we detect the Ba $4d$ and Cu $3p$ core levels around 90 and 75 eV of binding energy ($E_B$), respectively. As expected, the spectrum of BaCu$_2$As$_2$ exhibits additional peaks around 41 eV corresponding to the As $3d$ states that do not appear in the spectrum of $\alpha$-BaCu$_2$Sb$_2$. However, 4 features instead of 2 are observed, suggesting a surface state similar to the one reported previously for the EuFe$_2$As$_{2-x}$P$_x$ system \cite{richard2014JPCM}. The spectra of $\alpha$-BaCu$_2$Sb$_2$ exhibits 4 peaks around $E_B=32$ eV assigned to the Sb $4d$ electronic states. In this case though, the existence of four Sb peaks does not indicate the presence of a surface state since the crystal structure of $\alpha$-BaCu$_2$Sb$_2$ itself contains 2 inequivalent Sb sites.  

A rigid band shift is the simplest assumption one can make when describing the effect of doping. When considering only the \emph{shape} of the quasi-particle dispersions, this assumption works surprisingly well in the (Ba,K)Fe$_2$As$_2$ and Ba(Fe,Co)$_2$As$_2$ systems \cite{Neupane_PRB2011}, despite a strong dependence of the strength of correlations and coherence on the filling \cite{WernerNatPhys8}. It is also still valid when the filling of the $3d$-shell reaches 7 electrons in BaCo$_2$As$_2$ \cite{xu2013PRX}. As illustrated in the left inset of Fig. \ref{Fig1_core}(d) that compares the As $3d$ core levels of BaFe$_2$As$_2$, BaCo$_2$As$_2$ and BaCu$_2$As$_2$, the As$3d$ core levels in BaCo$_2$As$_2$ are downshifted in energy compared to their energy position in BaFe$_2$As$_2$, indicating an upward shift of the chemical potential. Assuming that the Cu $3d$ shell in BaCu$_2$As$_2$ contains 9 electrons as in the cuprates, a large downshift of the core level position should be observed. In contrast, the center of gravity of the As $3d$ core levels in BaCu$_2$As$_2$ is found nearly at the same energy as in BaFe$_2$As$_2$. This indicates clearly that the rigid band shift approximation is no longer valid and that other effects, such as the electronic valency of As, must be considered. A direct corollary is that the valency of the Cu atoms, which have As as ligands, may also be strongly affected.

\begin{figure}[!t]
\begin{center}
\includegraphics[width=3.4in]{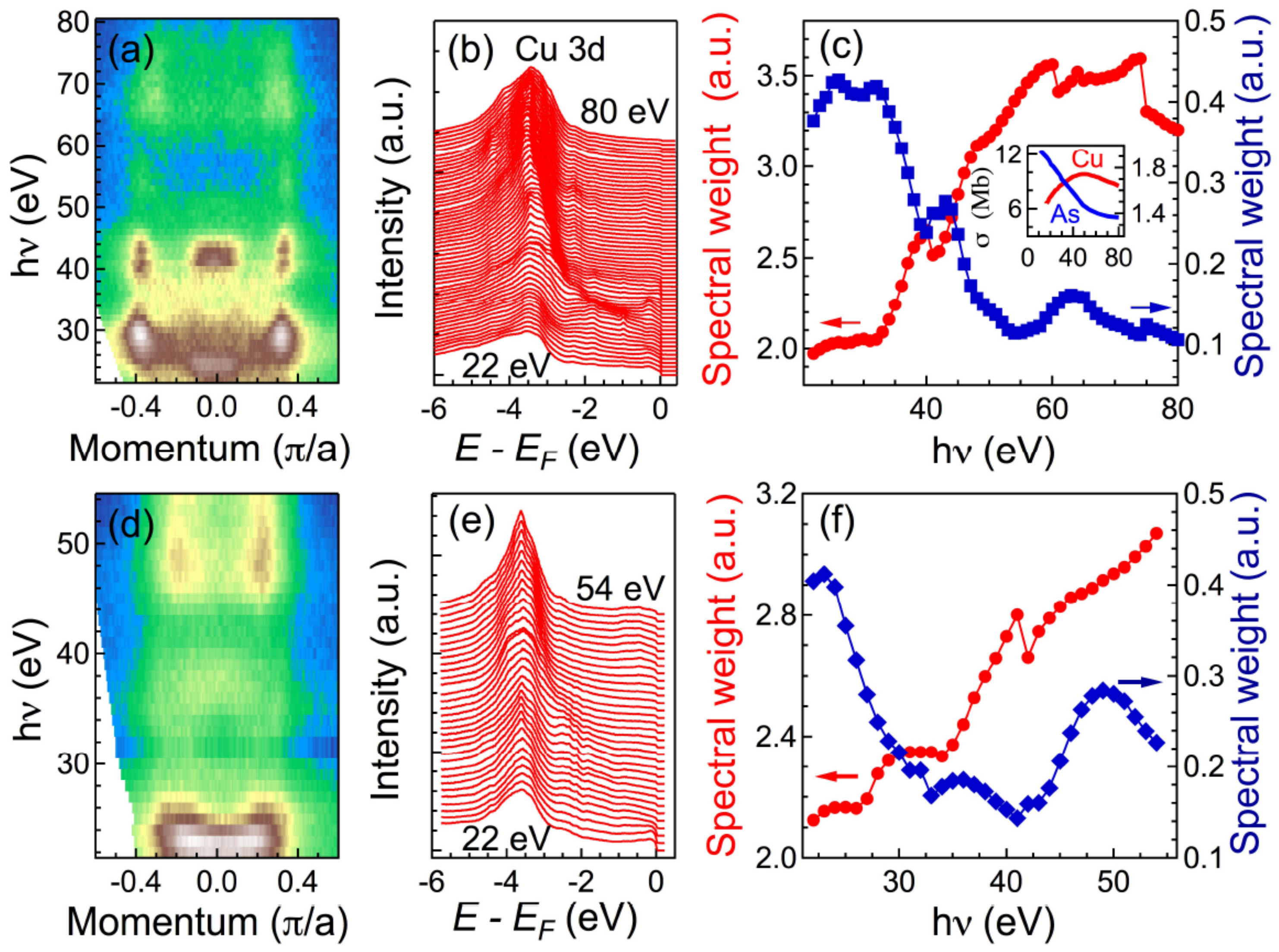}
\end{center}
\caption{\label{Fig2_kz}(Color online) (a) Photon energy ($h\nu$) dependence of the ARPES intensity (integration within $E_F\pm10$ meV) in BaCu$_2$As$_2$ recorded at 20 K along $\Gamma$-M. (b) Corresponding EDCs curves. (c) $h\nu$ dependence of the spectral weight in BaCu$_2$As$_2$ integrated within the [-4, -3] eV (red) and [-1, 0] eV (blue) energy ranges. The inset shows the Cu $3d$ and As $4p$ photoemission cross sections. (d)-(f) Same as (a)-(c) but for $\alpha$-BaCu$_2$Sb$_2$.}
\end{figure}

To check this latter scenario, we performed ARPES measurements of the valence states. In Fig. \ref{Fig2_kz}(a) we display the photon energy ($h\nu$) dependence of the ARPES intensity at $E_F$ recorded along $\Gamma$-M in BaCu$_2$As$_2$. Although the intensity is modulated along $k_z$, the momentum position of the features observed varies very little with $h\nu$, suggesting a quasi-two-dimensional electronic structure. Thus, we can tentatively assign the $h\nu$ values 41 eV and 36 eV to $k_z=0$ and $k_z=\pi$, but we cannot unambiguously determine which one is which. Despite a different crystal structure, very similar results are observed for $\alpha$-BaCu$_2$Sb$_2$, as shown in Fig. \ref{Fig2_kz}(d), except that the $k_z$ values associated with $h\nu=36$ eV and 41 eV are exchanged. 

The $h\nu$ dependence of the normal emission energy distribution curve (EDC) in BaCu$_2$As$_2$ and $\alpha$-BaCu$_2$Sb$_2$ are displayed in Figs. \ref{Fig2_kz}(b) and \ref{Fig2_kz}(e), respectively. Unlike the ferropnictide materials, for which strong spectral weight associated to the Fe $3d$ states is observed near $E_F$ \cite{Ding_JPCM2011}, BaCu$_2$As$_2$ and $\alpha$-BaCu$_2$Sb$_2$ exhibit only very small intensity at $E_F$. On the other hand, the Cu-pnictides show a very strong peak around 3.5 eV below $E_F$. In Fig. \ref{Fig2_kz}(c) we compare the spectral intensity of the normal emission EDCs integrated in the [-4, -3] eV and [-1, 0] energy ranges. The $h\nu$ dependences of these spectral intensities show different trends, thus indicating that their origin is different. While the spectral intensity in the [-4, -3] eV range increases with $h\nu$ increasing from 22 eV to about 60 eV and then slowly decreases, the spectral intensity near $E_F$ is at its highest at low $h\nu$ values and drops with $h\nu$ increasing. Besides additional features, like the small peaks around 41 eV and 64 eV in the [-1, 0] energy range that are likely induced by the $k_z$ effect on the photoemission matrix elements, the intensity of the [-4, -3] eV and [-1, 0] eV ranges are qualitatively in reasonable agreement with the photoemission cross sections \cite{yeh1985cross_section} of the Cu $3d$ and As $4p$ states, respectively, which are reproduced in the inset of \ref{Fig2_kz}(c). A similar observation is made for $\alpha$-BaCu$_2$Sb$_2$, as illustrated by Figs. \ref{Fig2_kz}(e)-(f).

\begin{figure}[!t]
\begin{center}
\includegraphics[width=3.4in]{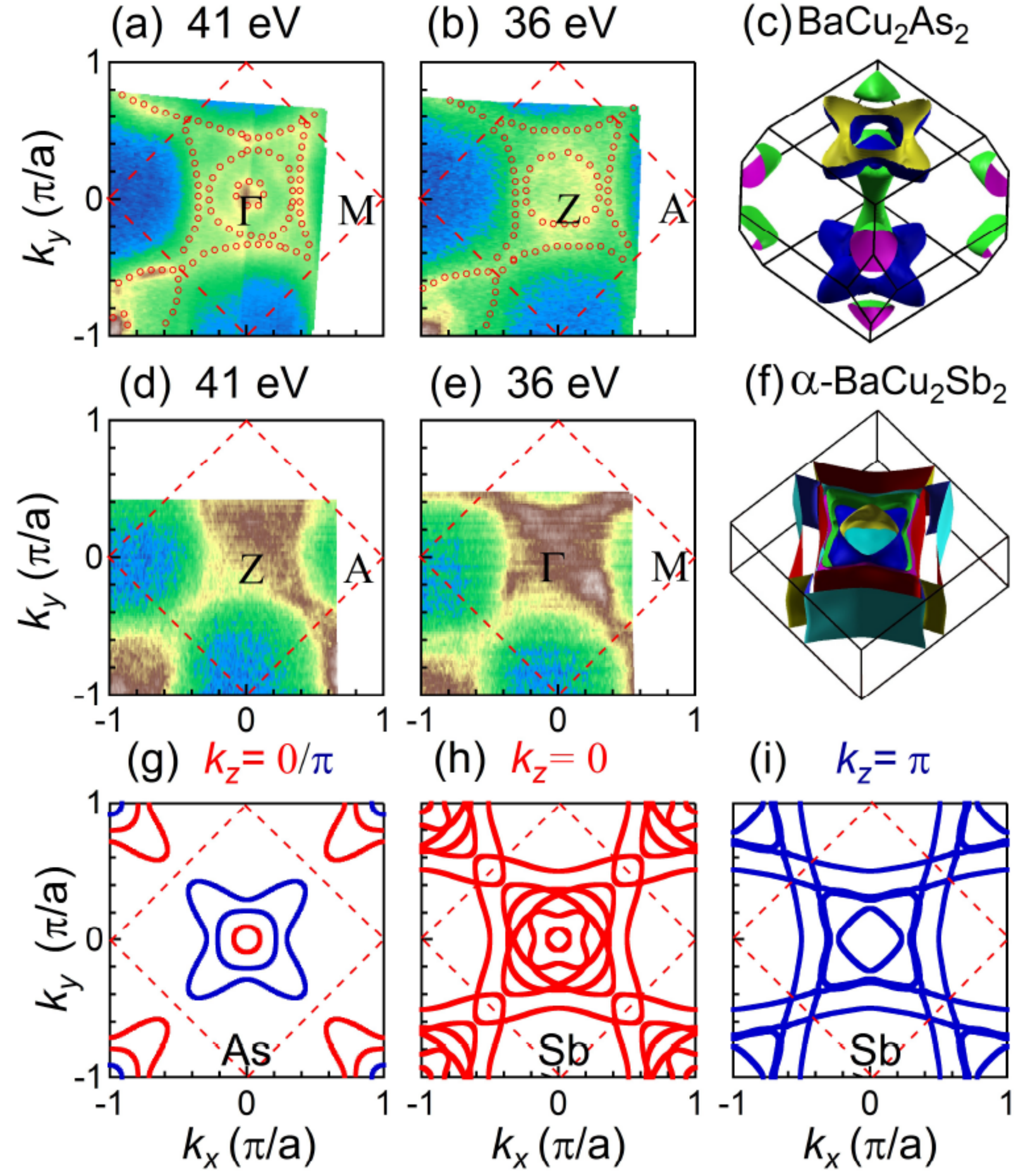}
\end{center}
\caption{\label{Fig3_FS}(Color online) (a) and (b) ARPES intensify plot of BaCu$_2$As$_2$ recorded at 20 K ($\pm 10$ meV integration around $E_F$) with 41 eV and 36 eV, respectively. The dots are guides for the eye for the Fermi surfaces. (c) Calculated three-dimensional Fermi surface of BaCu$_2$As$_2$. (d) and (e) Same as (a) and (b) but for $\alpha$-BaCu$_2$Sb$_2$. (f) Calculated three-dimensional Fermi surface of $\alpha$-BaCu$_2$Sb$_2$. (g) Calculated GGA Fermi surface of BaCu$_2$As$_2$ at $k_z=0$ (red) and $k_z=\pi$ (blue). (h) Calculated GGA Fermi surface of $\alpha$-BaCu$_2$Sb$_2$ at $k_z=0$. (i) Calculated GGA Fermi surface of $\alpha$-BaCu$_2$Sb$_2$ at $k_z=\pi$.}
\end{figure}

\begin{figure*}[!t]
\begin{center}
\includegraphics[width=7in]{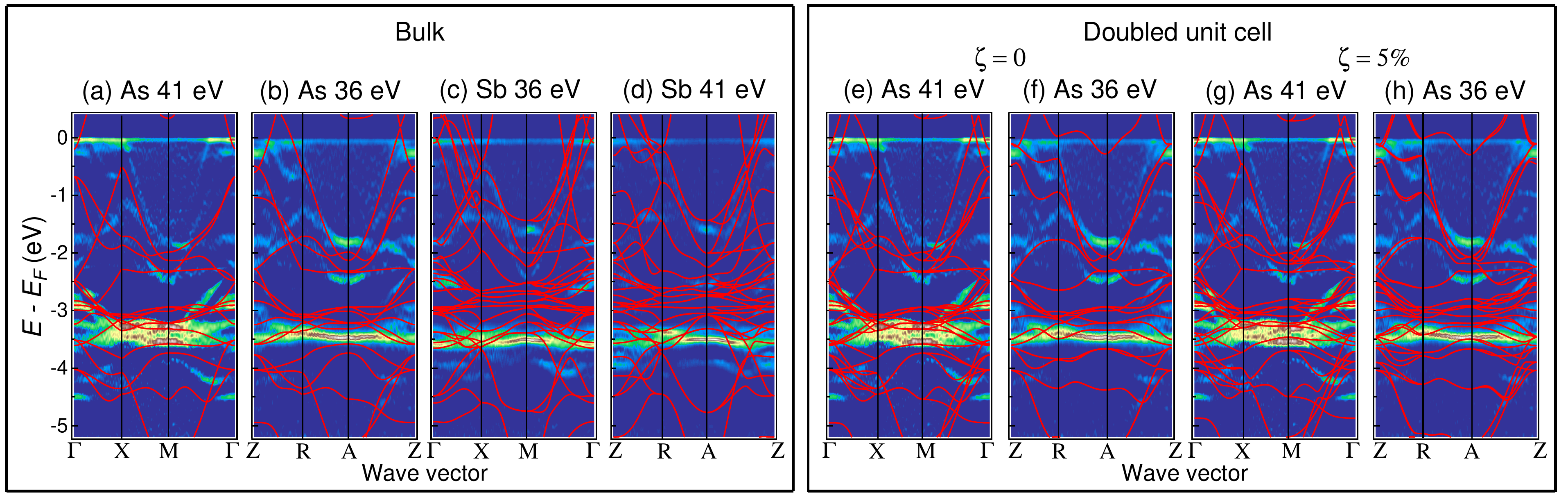}
\end{center}
\caption{\label{Fig4_LDA}(Color online) Comparison of unrenormalized GGA calculations (red curves) and curvature intensity plots for the electronic dispersion along high symmetry lines: (a) BaCu$_2$As$_2$ at 41 eV, GGA at $k_z=0$. (b) BaCu$_2$As$_2$ at 36 eV, GGA at $k_z=\pi$. (c) $\alpha$-BaCu$_2$Sb$_2$ at 36 eV, GGA at $k_z=0$. (d) $\alpha$-BaCu$_2$Sb$_2$ at 41 eV, GGA at $k_z=\pi$. (e)-(f) Same as (a) and (b) but with the calculations done for a hypothetical BaCu$_2$As$_2$ compound where we have lowered the crystal symmetry as if two consecutive CuAs layers were inequivalent, in order to simulate the surface. The relative variation of the As height position with respect to the Cu plane, for the two consecutive As layers, $\zeta$, is fixed to 0. (g)-(h) Same as in (e) and (f) but for the parameter $\zeta=5$\%.}
\end{figure*}

The results discussed above indicate clearly that most of the Cu $3d$ states are located at about 3.5 eV below $E_F$. We thus conclude that unlike in the cuprates, the $3d$ shell of Cu is completely filled, an experimental conclusion consistent with a previous prediction based on a local density approximation (LDA) study of BaCu$_2$As$_2$ and SrCu$_2$As$_2$ \cite{singh2009electronic}. Experimentally, a Fermi surface different from that of the ferropnictides, essentially made of pnictogen states, is thus expected. In Figs. \ref{Fig3_FS}(a) and \ref{Fig3_FS}(b), we compare the Fermi surface intensity maps obtained on BaCu$_2$As$_2$ using $h\nu=41$ eV and 36 eV, respectively. In agreement with the $h\nu$ dependence of the intensity plot shown in \ref{Fig2_kz}(a) and despite a redistribution of spectral intensity along the Fermi surface, the measured Fermi surfaces for these two set of data are surprisingly similar, indicating a quasi-two-dimensional electronic structure. This is in sharp contrast with our GGA calculation of the three-dimensional Fermi surface of this material, which are displayed in Fig. \ref{Fig3_FS}(c), as well as with the calculated Fermi surface cuts at $k_z=0$ and $k_z=\pi$, shown in Fig. \ref{Fig3_FS}(g). Even more surprising is the strong resemblance between the results recorded on BaCu$_2$As$_2$, and those recorded on $\alpha$-BaCu$_2$Sb$_2$, which are displayed in Figs. \ref{Fig3_FS}(d) and \ref{Fig3_FS}(e), although the features obtained for $\alpha$-BaCu$_2$Sb$_2$ are broader, which we attribute to a bad surface quality due to the difficulty of cleaving this material with strong Cu-Sb inter-layer bonding. Unlike BaCu$_2$As$_2$, $\alpha$-BaCu$_2$Sb$_2$ is predicted to have a two-dimensional Fermi surface, as illustrated in Fig. \ref{Fig3_FS}(f) and by the Fermi surface cuts at $k_z=0$ and $k_z=\pi$ given in Figs. \ref{Fig3_FS}(h) and \ref{Fig3_FS}(i), respectively. 

In Fig. \ref{Fig4_LDA}, we display the ARPES curvature \cite{zhang2011precise} intensity plots of BaCu$_2$As$_2$ and $\alpha$-BaCu$_2$Sb$_2$ recorded at 20 K along some high-symmetry lines with 41 eV and 36 eV photons. As mentioned above, the quasi-two-dimensional nature of the experimental Fermi surfaces prevents us from identifying unambiguously which one corresponds to $k_z=0$ and which one corresponds to $k_z=\pi$, but our $h\nu$ dependence data suggest that the roles are exchanged in BaCu$_2$As$_2$ and $\alpha$-BaCu$_2$Sb$_2$. For these plots we tentatively assume that 41 eV corresponds to $k_z=0$ ($k_z=\pi$) in BaCu$_2$As$_2$ ($\alpha$-BaCu$_2$Sb$_2$). Except for a few obvious discrepancies, like the absence of electron pocket at $\Gamma$ in the experimental data and the position of the hole-like dispersion around $\Gamma$, the GGA calculations at $k_z=0$ capture well the main features observed experimentally, without any renormalization. A similar comment can be made for $\alpha$-BaCu$_2$Sb$_2$, although in that case the agreement of the experimental data and the GGA predictions for the location of the Cu $3d$ bands is slightly worse. 

Given the indications of the presence of a surface state, we have tried to understand the striking resemblance of the measured BaCu$_2$As$_2$ spectral function with the one of $\alpha$-BaCu$_2$Sb$_2$, by assuming that the ARPES spectra display a substantial contribution from the surface layer. In BaFe$_2$As$_2$, a surface relaxation that consists essentially in an elongation of the As-height at the surface has been studied in Ref. \cite{Nascimento_PRL2009}. The result of such a structural surface relaxation is that the two FeAs layers of the surface unit cell are no longer equivalent, putting the surface crystal structure in the same symmetry group as the bulk of $\alpha$-BaCu$_2$Sb$_2$ -- namely P4/nmm -- where two inequivalent layers appear due to the inversion of Cu and Sb. To the best of  our knowledge, the precise surface crystal structure of BaCu$_2$As$_2$ has not been characterized yet, but it seems plausible that the ``collapsed" character of this phase rather enhances such a structural distortion at the surface, and it is safe to postulate that a substantial distortion, leaving the two upper-most CuAs surface layers inequivalent, occurs. 

Now, we can speculate that chemically, substituting Sb by As does not result in a drastic difference in the spectra, as long as the crystal symmetry is preserved. We have checked this hypothesis by calculating the band structure of a hypothetical BaCu$_2$As$_2$ compound where we have lowered the crystal symmetry as if the two layers were inequivalent. The resulting bands are overlaid to the experimental data in Figs. \ref{Fig4_LDA}e and \ref{Fig4_LDA}f. We also show the band structure that we obtain if we increase the As height by 5\% on one of the layers, in Figs. \ref{Fig4_LDA}g and \ref{Fig4_LDA}h. One of the consequences of such a doubled unit cell is that the number of bands in the $k_z=0$ plane is multiplied by two due to folding (Figs. \ref{Fig4_LDA}e and \ref{Fig4_LDA}g), and the dispersion corresponds to a superposition of the bands in the $k_z=0$ and $k_z=\pi$ plane of the I4/mmm structure (Figs. \ref{Fig4_LDA}a and \ref{Fig4_LDA}b). In the $k_z=\pi$ plane of the doubled unit cell, the number of bands is also multiplied by two but they are degenerate if the two layers have the same structure (Fig. \ref{Fig4_LDA}f), while a difference appears if we introduce a distortion (Fig. \ref{Fig4_LDA}h). While the knowledge of the precise surface structure would be needed to obtain quantitative agreement, Figs. \ref{Fig4_LDA}e-\ref{Fig4_LDA}h strongly suggest that the similarity of the BaCu$_2$As$_2$ spectra to the $\alpha$-BaCu$_2$Sb$_2$ ones can indeed be attributed to a lowering of the surface crystal symmetry of the kind we describe.

Finally, we comment on the 3$d^{10}$ configuration of Cu. Singh already pointed out \cite{singh2009electronic} that the much shorter $c$ parameter in BaCu$_2$As$_2$ as compared to BaNi$_2$As$_2$ and BaFe$_2$As$_2$ is suggestive of different bonding in BaCu$_2$As$_2$, which is confirmed by our As $3d$ core level spectra. In fact, comparison of the lattice parameters places BaCu$_2$As$_2$ in the collapsed tetragonal phase \cite{anand2012crystal}, which is associated with a stronger As-As interlayer bonding \cite{Hoffmann_JPC89,YildirimPRL102}. However, our results suggest that the strong As-As interlayer bonding cannot be the cause of the 3$d^{10}$ configuration of Cu in BaCu$_2$As$_2$. Indeed, the 3$d$ Cu states in $\alpha$-BaCu$_2$Sb$_2$ are located practically at the same energy as in BaCu$_2$As$_2$, despite the absence of Sb-Sb direct interlayer bonding in the former material, thus reinforcing previous arguments by Anand \emph{et al.} based on the resemblance of the physical properties of SrCu$_2$As$_2$ and $\alpha$-SrCu$_2$Sb$_2$ \cite{anand2012crystal}. Interestingly, the Cu dopant states in Cu-substituted BaFe$_2$As$_2$ are also found around 3-4 eV below $E_F$ \cite{mcleod2012effectXAS,ideta2013dependence}, suggesting that Cu-dopants are already in a +1 oxidation state. This is consistent with the absence of any shift in the Cu $2p$ core level spectra as a function of doping \cite{YJ_Yan_PRB87}, as well as with the suppression of magnetic moment and the observation of superconductivity only for a very narrow Cu substitution range \cite{Mun_PRB80,ni2010temperature}, in contrast to substitution of Fe by Co \cite{Sefat_PRL2008,N_Ni_PRB78} and Ni \cite{Li_NJP2009}. However, the clear observation of added electron carriers in Ba(Fe$_{1-x}$Cu$_x$)$_x$As$_2$ \cite{Mun_PRB80,ni2010temperature} indicates that the Cu substitution is electron doping nevertheless, suggesting that the local As-As and As-transition metal bondings are strongly affected, a situation that could be stabilized at high substitution levels by the collapsed tetragonal phase. Our findings support the hypothesis \cite{anand2012crystal} that it is the stability of the 3$d^{10}$ configuration of Cu (Cu$^{+1}$) that favors the collapsed tetragonal phase of BaCu$_2$As$_2$, rather than the other way around.

In summary, we have performed angle-resolved photoemission spectroscopy measurements on BaCu$_2$As$_2$ and $\alpha$-BaCu$_2$Sb$_2$ to extract their electronic band dispersions and Fermi surfaces. We found that most of the Cu 3$d$ spectral weight locates around 3-4 eV below $E_F$, whereas the intensity around  $E_F$ mainly comes from As 4$p$ bands, suggesting a filled Cu $3d$ shell. The observation of split As $3d$ core levels and the absence of pronounced three-dimensionality in the measured electronic structure of BaCu$_2$As$_2$ is compatible with a surface state emerging from the cleavage of this material in the collapsed tetragonal phase. However, the observation of similar Cu $3d$ states in $\alpha$-BaCu$_2$Sb$_2$ without pnictide-pnictide interlayer bonding suggests that the stability of the Cu$^{+1}$ configuration favors the collapsed tetragonal phase rather than the other way around. Our study indicates that BaCu$_2$As$_2$ and $\alpha$-BaCu$_2$Sb$_2$ are $sp$ metals with weak electronic correlations.

We acknowledge W.-L. Zhang for useful discussions. This work was supported by grants from MOST (2010CB923000 and 2011CBA001000, 2011CBA00102, 2012CB821403) and NSFC (10974175, 11004232, 11034011/A0402, 11234014 and 11274362) from China, the Cai Yuanpei program, the French ANR via project PNICTIDES, IDRIS/GENCI under project 091393 and the European Research Council under Project No. 617196. This work is based in part on research conducted at the Synchrotron Radiation Center, which was primarily funded by the University of Wisconsin-Madison with supplemental support from facility Users and the University of Wisconsin-Milwaukee. The work at ORNL was supported by the Department of Energy, Basic Energy Sciences, Materials Sciences and Engineering Division.

\bibliography{biblio_long}

\end{document}